\documentclass[aps,prl,reprint,superscriptaddress,showpacs,amsmath,amssymb]{revtex4-1}

\usepackage{graphicx}% Include figure files

\begin{document}

% Use the \preprint command to place your local institutional report
% number in the upper righthand corner of the title page in preprint mode.
% Multiple \preprint commands are allowed.
% Use the 'preprintnumbers' class option to override journal defaults
% to display numbers if necessary
%\preprint{}

%Title of paper
\title{Field-Induced Lifshitz Transition without Metamagnetism in CeIrIn\(_5\)}

% repeat the \author .. \affiliation  etc. as needed
% \email, \thanks, \homepage, \altaffiliation all apply to the current
% author. Explanatory text should go in the []'s, actual e-mail
% address or url should go in the {}'s for \email and \homepage.
% Please use the appropriate macro foreach each type of information

% \affiliation command applies to all authors since the last
% \affiliation command. The \affiliation command should follow the
% other information
% \affiliation can be followed by \email, \homepage, \thanks as well.

\author{D. Aoki}
\affiliation{Institute for Materials Research, Tohoku University, Oarai, Ibaraki, 311-1313, Japan}
\affiliation{Univ. Grenoble Alpes, INAC-SPSMS, F-38000 Grenoble, France}
\affiliation{CEA, INAC-SPSMS, F-38000 Grenoble, France}

\author{G. Seyfarth}
\affiliation{Univ. Grenoble Alpes, LNCMI, 38042 Grenoble, France}
\affiliation{Laboratoire National des Champs Magn\'{e}etiques Intenses (LNCMI-EMFL), CNRS, UJF, 38042 Grenoble, France}

\author{A. Pourret}
\affiliation{Univ. Grenoble Alpes, INAC-SPSMS, F-38000 Grenoble, France}
\affiliation{CEA, INAC-SPSMS, F-38000 Grenoble, France}

\author{A. Gourgout}
\affiliation{Univ. Grenoble Alpes, INAC-SPSMS, F-38000 Grenoble, France}
\affiliation{CEA, INAC-SPSMS, F-38000 Grenoble, France}

\author{A. McCollam}
\affiliation{High Field Magnet Laboratory (HFML-EMFL), Radboud University, 6525 ED Nijmegen, The Netherlands}

\author{J.A.N. Bruin}
\affiliation{High Field Magnet Laboratory (HFML-EMFL), Radboud University, 6525 ED Nijmegen, The Netherlands}

\author{Y. Krupko}
\affiliation{Laboratoire National des Champs Magn\'{e}etiques Intenses (LNCMI-EMFL), CNRS, UJF, 38042 Grenoble, France}

\author{I. Sheikin}
\email[]{ilya.sheikin@lncmi.cnrs.fr}
%\homepage[]{Your web page}
%\thanks{}
%\altaffiliation{}
\affiliation{Laboratoire National des Champs Magn\'{e}etiques Intenses (LNCMI-EMFL), CNRS, UJF, 38042 Grenoble, France}

\date{\today}

\begin{abstract}
We report high magnetic field measurements of magnetic torque, thermoelectric power (TEP), magnetization, and de Haas-van Alphen (dHvA) effect in CeIrIn\(_5\) across 28 T, where a metamagnetic transition was suggested in previous studies. The TEP displays two maxima at 28 T and 32 T. Above 28 T, a new, low dHvA frequency with strongly enhanced effective mass emerges, while the highest frequency observed at low field disappears entirely. This suggests a field-induced Lifshitz transition. However, longitudinal magnetization does not show any anomaly up to 33 T, thus ruling out a metamagnetic transition at 28 T.
\end{abstract}

% insert suggested PACS numbers in braces on next line
% insert suggested keywords - APS authors don't need to do this
%\keywords{}

%\maketitle must follow title, authors, abstract, \pacs, and \keywords
\maketitle

% body of paper here - Use proper section commands
% References should be done using the \cite, \ref, and \label commands

%\section{Introduction}

An electronic topological transition, better known as a Lifshitz transition (LT), is a change of the Fermi surface (FS) topology of a metal due to variation of the Fermi energy and/or the band structure~\cite{Lifshitz1960}. Possible methods of changing the band structure and the relative position of the Fermi energy within the band structure are, for example, alloying or the application of external pressure. In some cases, a topological change of the FS can also be induced by magnetic field, due to the Zeeman splitting of the electronic bands. Contrary to most conventional phase transitions, the LT is not associated with any symmetry breaking. Additionally, the LT is a quantum phase transition as it is a continuous transition only at $T = 0$; it becomes a crossover at finite temperatures.

The LT has recently come to prominence in modern solid state physics. A LT was argued to occur in iron pnictides~\cite{Liu2010,Khan2014}, high temperature superconductors~\cite{Norman2010,LeBoeuf2011} and the strongly correlated electron system Na$_x$CoO$_2$~\cite{Okamoto2010}. In all these materials the LT was induced by either doping or hydrostatic pressure. Heavy fermion (HF) compounds, on the other hand, appear to be good candidates for LTs induced by magnetic field. Indeed, the Zeeman splitting of the narrow electronic bands crossing the Fermi level can readily induce topological changes of the FS in such materials.

The subject of field-induced LTs in HF materials has already received a thorough theoretical treatment within different models~\cite{Schlottmann2011, Bercx2012, Benlagra2013, Burdin2013}. However, there is still a lack of experimental data mostly because magnetic fields higher than those available in most laboratories are often required to induce a LT for HF. One notable exception is CeRu\(_2\)Si\(_2\), where a field induced metamagnetic transition takes place at about 8 T applied along the \(c\)-axis. Daou \emph{et al}~\cite{Daou2006} demonstrated that this transition is accompanied by a continuous evolution of the FS, where one of the spin-split sheets of the heaviest surface shrinks to a point. Another example is YbRh\(_2\)Si\(_2\)~\cite{Rourke2008} where a low-field ``large" FS, including the Yb 4\(f\)-quasihole, is increasingly spin-split until a majority-spin branch undergoes a LT and disappears at a metamagnetic transition that occurs at about 10 T. More recent TEP measurements~\cite{Pfau2013,Pourret2013} detected three successive field-induced transitions in YbRh\(_2\)Si\(_2\), which were identified as LTs by renormalized band structure calculations~\cite{Pfau2013}.

CeIrIn\(_5\) is a non-magnetic HF superconductor with \(T_c = 400\) mK~\cite{Petrovic2001}. It crystallizes in the tetragonal HoCoGa\(_5\) structure (space group \(P4/mmm\)). A large electronic specific heat coefficient \(\gamma = 750 \; \rm{mJ/K^2mol}\)~\cite{Movshovich2001} suggests strongly enhanced effective masses. Indeed, effective masses of up to $\approx30 m_0$ were directly observed in dHvA measurements performed in magnetic field up to 17 T~\cite{Haga2001}. These measurements together with band structure calculations revealed two quasi-two-dimensional \(\alpha\) (electron) and \(\beta\) (hole) FS sheets with itinerant \(4f\)-electrons.

Previous magnetic torque measurements at 45 mK on CeIrIn\(_5\) revealed a kink slightly below 30 T for magnetic field applied along the \(c\)-axis, which was interpreted as a metamagnetic transition~\cite{Palm2003}. However, pulsed field magnetization measurements performed at $T = 1.3$ K suggest a weak metamagnetic transition at a much higher field of 42 T~\cite{Takeuchi2001}. In addition, a field-induced transition above 30 T was observed in specific heat measurements down to 1.6 K~\cite{Kim2002}. The extrapolation of the transitions observed in specific heat to zero temperature yields 26 T, a value comparable to the field where an anomaly in magnetic torque was observed~\cite{Palm2003}. More recently, the results of resistivity and dHvA measurements by torque magnetometery up to 45 T were reported~\cite{Capan2009}: a kink in magnetic torque was observed at 28 T in agreement with previous results~\cite{Palm2003}. At this field, the resistivity exhibits a broad maximum with a subsequent decrease. The dHvA measurements revealed a small change of the dHvA frequencies across the tentative metamagnetic transition with no significant variation of the effective masses. The main observation was a strong damping of the oscillations above the anomaly.

In this paper, we report high field magnetic torque, magnetization, TEP and high resolution dHvA measurements on high quality single crystals of CeIrIn\(_5\). The field dependence of TEP shows a maximum at 28 T for field along the $c$-axis.  Above this field, dHvA measurements reveal a new low dHvA frequency with high effective mass, which is not observed at lower field. Furthermore, the highest dHvA frequency observed at low field disappears entirely at precisely the same field. We argue that these observations are most naturally accounted for by a continuous LT. Remarkably, the LT here is not associated with metamagnetism as no anomaly was observed in magnetization measurements in a strong contrast to CeRu\(_2\)Si\(_2\) and YbRh\(_2\)Si\(_2\).

High quality single crystals of CeIrIn\(_{5}\) used in our studies were grown by an In self-flux method. dHvA measurements were performed by a conventional torque magnetometry technique. This was done using either a metallic cantilever in a top-loading dilution refrigerator in field up to 34 T or a microcantilever in a $^3$He cryostat up to 33 T. In the latter case, the magnetization of the same sample was extracted by subtracting zero gradient data from the curve obtained in the presence of a strong field gradient~\cite{McCollam2011}. TEP measurements were done using a ``one heater, two thermometers" set-up in a $^3$He cryostat down to 480~mK and up to 34~T. Details of the TEP measurements are described elsewhere~\cite{Boukahil2014}.

\begin{figure}[htb]
\includegraphics[width=7.5cm]{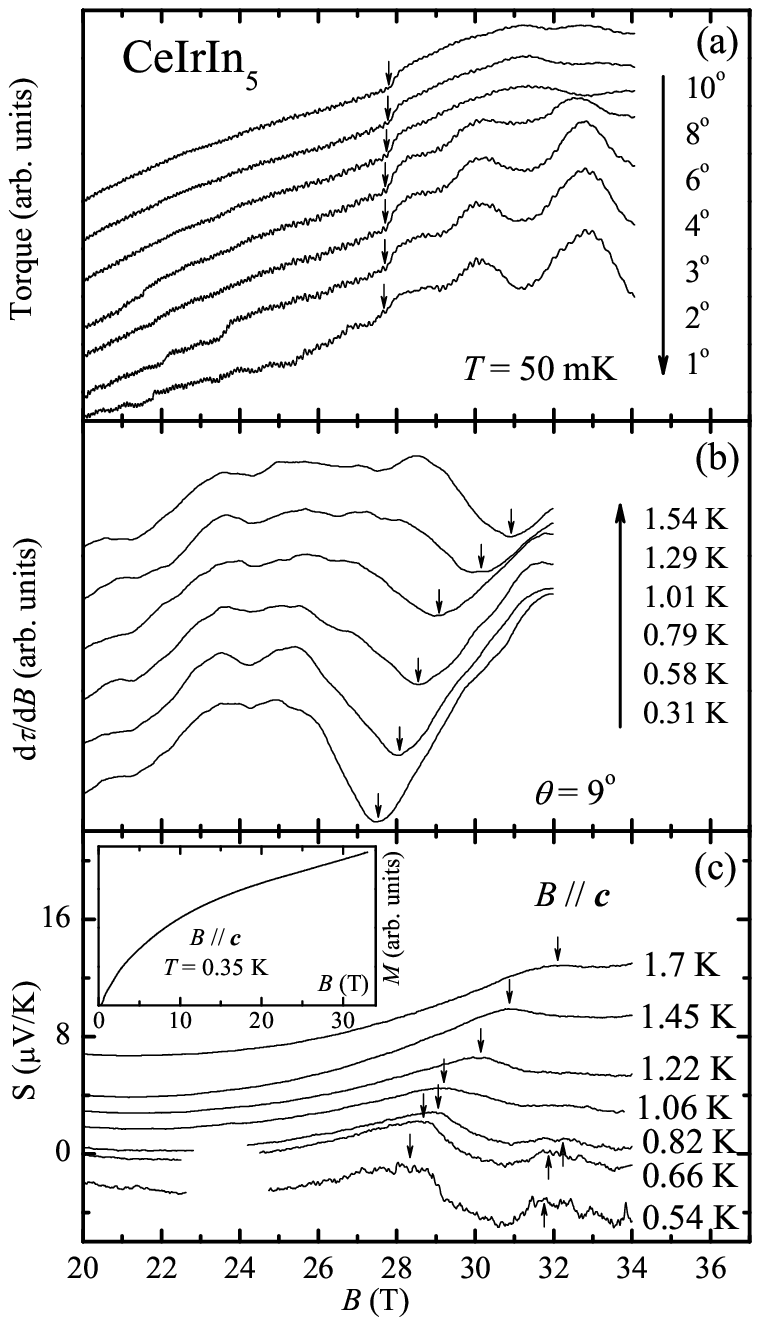}
\caption{\label{fig:Field_dependence}(a) Low temperature magnetic torque \(\tau\), as a function of field \(B\), applied at different angles from the \(c\)-axis. Lines are vertically shifted for clarity. (b) Derivative of torque signal vs magnetic field applied at 9\(^\circ\) from the \(c\)-axis at different temperatures. Lines are shifted vertically for clarity. (c) TEP \(S\) as a function of \(B\) applied along the \(c\)-axis at different temperatures. The inset shows longitudinal magnetization \(M\) vs \(B\) at 0.35 K.}
\end{figure}

Fig.~\ref{fig:Field_dependence} shows the field dependence of magnetic torque, $\tau$, at different field orientations and temperatures as well as TEP, also known as the Seebeck coefficient, $S$, with field along the $c$-axis. In agreement with previous measurements~\cite{Palm2003,Capan2009}, $\tau(B)$ shows a distinct kink at $\approx28$ T at low temperature. At higher temperature, the kink becomes less clear, but the anomaly can be easily traced as a minimum in the field derivative of $\tau(B)$ (fig.~\ref{fig:Field_dependence}(b)). A new and particularly interesting result comes from the $S(B)$ dependence, shown for different temperatures in fig.~\ref{fig:Field_dependence}(c). At low temperature, the TEP shows a small but sharp positive peak at $\approx28$~T, which broadens when the temperature increases. This kind of anomaly was observed in both CeRu$_2$Si$_2$~\cite{Pfau2012,Boukahil2014} and YbRh$_2$Si$_2$~\cite{Pfau2013,Pourret2013} and is consistent with a reconstruction of the FS due to a LT of the polarized band. This anomaly occurs at about the same field as its counterpart in magnetic torque. A second anomaly in TEP, not observed in any previous measurements, occurs at higher field, around 32~T. This is not necessarily surprising as multiple anomalies were observed in TEP at the LT in both CeRu$_2$Si$_2$~\cite{Boukahil2014} and YbRh$_2$Si$_2$~\cite{Pfau2013,Pourret2013}, where all the transport and other thermodynamic measurements detected only one. Indeed, the presence of multiple anomalies seems to be a generic feature of the TEP at a LT in HF multiband systems.

\begin{figure}[htb]
%\centering
\includegraphics[width=7.5cm]{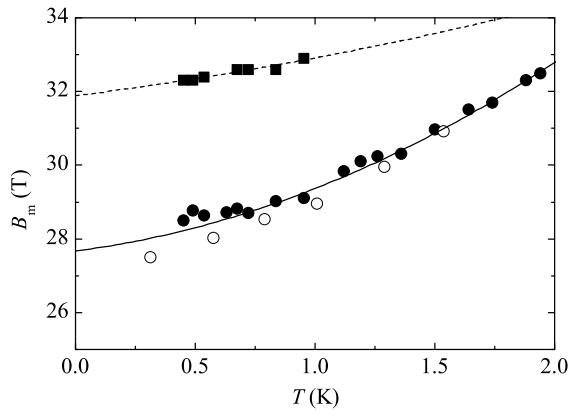}
\caption{High field phase diagram of CeIrIn$_5$ obtained from TEP (solid symbols) and magnetic torque measurements (open symbols). Symbols correspond to anomalies indicated by arrows in fig.~\ref{fig:Field_dependence}, whereas lines are guides for the eyes only.}
\label{fig:PhaseDiag.eps}
\end{figure}

In fig.~\ref{fig:PhaseDiag.eps}, we present the revision of the previously suggested high field magnetic phase diagram of CeIrIn$_5$~\cite{Capan2009}, based on the anomalies observed in TEP and magnetic torque signals. The lower field anomaly, observed both in TEP and torque measurements, is almost field-independent at low temperature in agreement with previous report~\cite{Capan2009}. The anomaly at higher field was observed in TEP only and not in any other measurements reported so far. It follows a similar temperature dependence as its lower field counterpart.

\begin{figure}[htb]
\includegraphics[width=7.5cm]{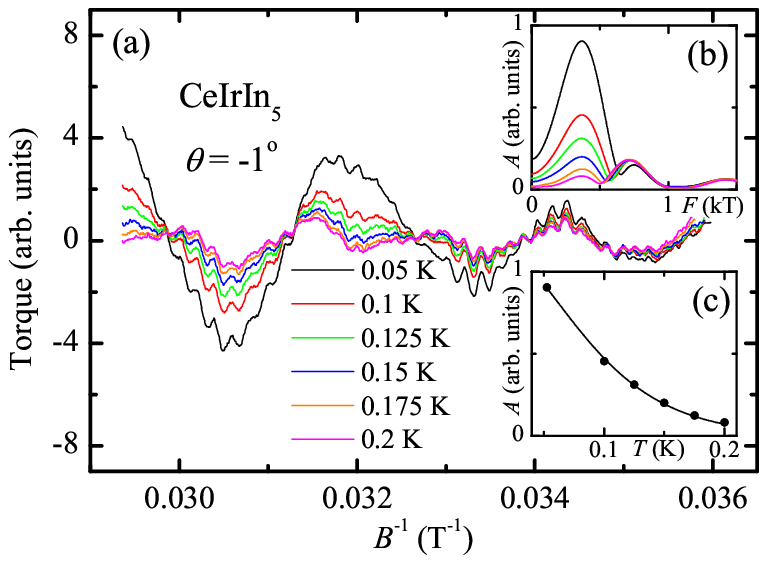}
\caption{\label{fig:lowF}(color online) (a) dHvA oscillations above 28 T for magnetic field applied at 1$^\circ$ from the $c$-axis at different temperatures. (b) Low frequency part of the Fourier spectra of oscillations from (a). (c) oscillatory amplitude as a function of temperature for the low frequency of 367 T, that emerges above 28 T. Line is a fit by the Lifshitz-Kosevich formula~\cite{Shoenberg2009}, yielding an effective mass of 54.1 \(m_0\).}
\end{figure}

Further evidence for a field-induced LT in CeIrIn$_5$ is provided by the analysis of the dHvA oscillations. In fig.~\ref{fig:Field_dependence}(a), one can clearly see a new low frequency that emerges above the transition even without subtracting the background. At $T = 50$ mK, the amplitude of these oscillations is much larger than that of the other frequencies. However, it rapidly decreases with temperature as shown in fig.~\ref{fig:lowF}(a). Fig.~\ref{fig:lowF}(b) shows the low frequency part of the fast Fourier  transform (FFT) spectrum for several temperatures that reveals a strong peak at 367 T. A fit of the oscillatory amplitude versus temperature by the temperature-dependent part of the Lifshitz-Kosevich formula~\cite{Shoenberg2009} shown in fig.~\ref{fig:lowF}(c) yields a huge effective mass of 54.1 $m_0$, which is much higher than all the masses observed at lower field~\cite{Haga2001}. Previous low field dHvA measurements also revealed a small frequency of 270 T, assigned as $\gamma$-branch~\cite{Haga2001}. However, the new frequency we observe above 28 T is distinct from the $\gamma$-branch, as we did not detect any low frequencies below that field. Furthermore, the amplitude of the oscillations originating from the $\gamma$-branch is extremely small, whereas the amplitude of the new low frequency oscillations is much stronger than that of the other frequencies. Finally, the reported effective mass of the $\gamma$-branch, 6.3 $m_0$, is the smallest among those observed at lower field, while the effective mass of the low frequency detected above 28 T is much higher than any of them. Thus, a very small but extremely heavy pocket of the FS emerges above the 28 T transition. Interestingly, an emergence of new dHvA frequencies with strongly enhanced effective masses was also observed in a sister compound CeCoIn$_5$ above 23 T~\cite{Sheikin2006}. However, in CeCoIn$_5$ all the low field dHvA frequencies are preserved above 23 T. This is not the case in CeIrIn$_5$, as we will discuss next.

\begin{figure}[htb]
%\centering
\includegraphics[width=7.5cm]{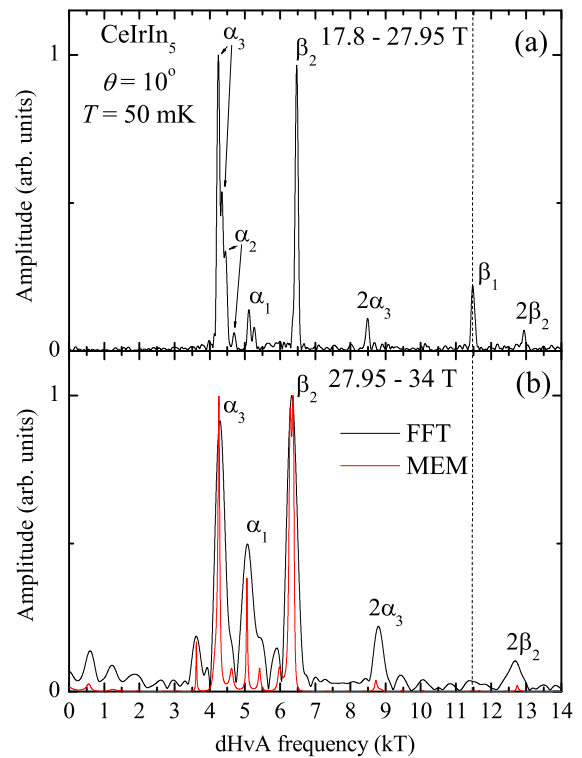}
\caption{(color online) Fourier spectra of the dHvA oscillations below (a) and above (b) the LT in CeIrIn\(_5\) with magnetic field at 10\(^\circ\) from the [001] direction. For the oscillations above the transition (b), both the spectra obtained from FFT (black) and MEM (red) are shown. The new low dHvA frequency that emerges above the transition was filtered out for clarity.}
\label{fig:HighF}
\end{figure}

Fig.~\ref{fig:HighF} shows FFTs of the quantum oscillations both below and above the 28 T transition for magnetic field applied at 10\(^\circ\) from [001], the largest field angle of our measurements. The FFT spectrum of the oscillations below the transition (fig.~\ref{fig:HighF}(a)) was obtained from a wide field range, 17.8 - 27.95 T. This allowed us to resolve all the fundamental frequencies, including the fine splitting of some of the $\alpha$-branches, in good agreement with previously reported lower field results~\cite{Haga2001}. Since the field range above the transition is limited, the FFT does not provide the same high frequency resolution. For this reason, we complimented our analysis of the oscillations above the transition by the Maximum Entropy Method (MEM), which was demonstrated to be a powerful technique for quantum oscillations analysis with a much higher frequency resolution as compared to FFT~\cite{Sigfusson1992}. Due to its remarkable stability to noise, MEM reveals oscillations with amplitudes even lower than the noise level~\footnote{MEM should, however, be used with caution as artificial peaks may appear on the resulting spectrum.}. The normalized spectra obtained from both the FFT and MEM are shown in fig.~\ref{fig:HighF}(b). Most of the fundamental frequencies observed below the transition are still present and are not significantly changed above it~\cite{[{There is a small continues variation of most of the dHvA frequencies with magnetic field, 1\% - 4\% over the investigated field range, due to magnetostriction. This drift accounts for a small shift of the second harmonic of the $\alpha_3$ frequency seen in Fig.~\ref{fig:HighF}. A similar variation of the dHvA frequencies was observed in CeRu$_2$Si$_2$: }][]{Aoki2001}}. The only notable exception is the highest \(\beta_1\)-frequency that disappears entirely above the transition, where neither the FFT nor the MEM analysis reveal any trace of it (fig.~\ref{fig:HighF}(b)). The same results were obtained at all the other orientations of the magnetic field. This is at variance with the previous high field dHvA study~\cite{Capan2009}, which reported only a strong damping of the oscillations originating from \(\beta_1\)- and \(\beta_2\)-branches above the transition~\footnote{In Ref.~\cite{Capan2009}, the observation of the \(\beta_1\)-frequency above the transition seems to be an artifact of the data analysis. Indeed, the \(\beta_1\)-frequency was observed for only one field orientation, for which the measurements were performed up to 33 T. The field dependence of the dHvA amplitudes was determined over equal inverse magnetic field intervals of 23 periods of $\alpha_3$. Thus, even the highest field range included some data from below the transition}.

Both the emergence of a new small heavy pocket of the FS and a complete disappearance of the highest dHvA frequency above 28 T in CeIrIn\(_5\) are most naturally accounted for by a field-induced LT. According to the band-structure calculations~\cite{Haga2001,Elgazzar2004,Choi2012}, \(\beta_1\) and \(\beta_2\) orbits originate from the same hole FS of band 14 and are centered around the $M$ and $A$ points in the momentum space respectively. While the \(\beta_2\) orbits are well separated from each other, the \(\beta_1\) orbits are almost touching at the $\Gamma$ point in the $\Gamma$-$X$-$M$ plane, as can be particularly well seen in Fig. 14 (b$''$) of Ref.~\cite{[][{. Here the FS of CeCoIn$_5$ is shown, but the FS of CeIrIn$_5$ is almost the same.}]Shishido2002}. Therefore, even a small expansion of this FS would lead to a topological transition, where the \(\beta_1\) orbit disappears, giving rise to a small pocket around the $\Gamma$ point, and, probably, another one around the $X$ point. This naive scenario naturally accounts for our findings. Field dependent band structure calculations are, however, required to confirm this scenario and to figure out how exactly the FSs reconstruct across the LT in CeIrIn$_5$.

We now discuss the differences between the LTs in CeIrIn\(_5\), and in CeRu$_2$Si$_2$ and YbRh$_2$Si$_2$. First of all, in both CeRu$_2$Si$_2$ and YbRh$_2$Si$_2$ the LT is accompanied by a clear anomaly in longitudinal magnetization. Indeed, magnetization shows a sharp step-like increase at about 8 T in CeRu$_2$Si$_2$~\cite{Haen1987,Flouquet2002}. In YbRh$_2$Si$_2$, the magnetization shows a kink rather than a step at 10 T and tends to saturate at higher field~\cite{Tokiwa2004,Tokiwa2005,Gegenwart2006} in agreement with theoretical predictions~\cite{ViolaKusminskiy2008}. On the contrary, in CeIrIn\(_5\) the magnetization curve is continuous without any anomaly at 28 T, as shown in the inset of fig.~\ref{fig:Field_dependence}(c). The only metamagnetic-like anomaly in CeIrIn\(_5\) was observed at a much higher field of 42 T in previous pulsed field measurements~\cite{Takeuchi2001}. Secondly, in both CeRu$_2$Si$_2$~\cite{Flouquet2002,Pfau2012} and YbRh$_2$Si$_2$~\cite{Tokiwa2004,Tokiwa2005,Gegenwart2006} a rapid decrease of the effective mass above the metamagnetic and LT field was observed, while there is a weak, if any, variation of the effective masses across the 28 T LT in CeIrIn\(_5\). Generally speaking, a LT is not expected to give rise to either a metamagnetic transition or a strong variation of the effective mass. In HF materials, however, it is believed that a magnetic field strong enough to sufficiently spin-split the electronic bands and induce a LT should produce a Zeeman energy comparable to a characteristic Kondo coherence energy, thus leading to a gradual destruction of the HF state and decrease of the effective mass. This is probably what happens in both CeRu$_2$Si$_2$ and YbRh$_2$Si$_2$. In CeIrIn\(_5\), measurements of the NMR Knight shift demonstrated an extreme robustness of the Kondo lattice coherence temperature and the effective mass, which hardly change at all even in magnetic field as high as 30 T~\cite{Shockley2013}. It is not clear at present whether CeIrIn$_5$ represents a unique case or a more general behavior of a field induced LT in HF systems, as our finding are difficult to reconcile with existing theories.

In conclusion, our dHvA measurements revealed the emergence of a new small but heavy pocket of the FS above 28 T along the $c$-axis in CeIrIn$_5$. Furthermore, the highest dHvA frequency, $\beta_1$, disappears completely at exactly the same field. These two observations represent an almost canonical case of a LT. This is further supported by the observation of two clear anomalies in TEP, at 28 and 32 T, typical for a LT. Most remarkably, the LT in this compound is not accompanied by any anomaly in the longitudinal magnetization. Rather, a kink in magnetic torque, implying a change of the transverse magnetization, was observed in our and previous measurements at the LT. Our results, therefore, call for a re-evaluation of the debate concerning explanations for the physics of CeIrIn$_5$ based on the presence of metamagnetism. In spite of the existence of numerous theoretical works dedicated to LTs in HF materials, further theoretical effort is required to explain our experimental findings in CeIrIn$_5$.

% If you have acknowledgments, this puts in the proper section head.
\begin{acknowledgments}
We acknowledge the support of the HFML-RU/FOM and the LNCMI-CNRS, members of the European Magnetic Field Laboratory (EMFL).
\end{acknowledgments}

% Create the reference section using BibTeX:
\bibliography{CeIrIn5}

\end{document}